\documentclass[aip,jcp,reprint,showpacs,floatfix,floats,nobalancelastpage,amsmath,amssymb]{revtex4-1}

\usepackage[dvips]{graphicx}  

\begin{document}


\title{Stochastic lumping analysis for linear kinetics and its application to the fluctuation relations between hierarchical kinetic networks}


\author{De-Ming Deng}
\affiliation{Institute of Physics, National Chiao Tung university, Hsinchu, 300, Taiwan}
\author{Cheng-Hung Chang} 
\affiliation{Institute of Physics, National Chiao Tung university, Hsinchu, 300, Taiwan}


\date{\today}

\begin{abstract}
Conventional studies of biomolecular behaviors rely largely on the construction of kinetic schemes. Since the selection of these networks is not unique, a concern is raised whether and under which conditions hierarchical schemes can reveal the same experimentally measured fluctuating behaviors and unique fluctuation related physical properties. To clarify these questions, we introduce stochasticity into the traditional lumping analysis, generalize it from rate equations to chemical master equations and stochastic differential equations, and extract the fluctuation relations between kinetically and thermodynamically equivalent networks under intrinsic and extrinsic noises. The results provide a theoretical basis for the legitimate use of low-dimensional models in the studies of macromolecular fluctuations and, more generally, for exploring stochastic features in different levels of contracted networks in chemical and biological kinetic systems.
\end{abstract}

\maketitle

\section{Introduction}
Kinetic schemes are widely used for studying the thermodynamic, dynamic, and stochastic
properties of macromolecules 
\cite{Jackson}.
These schemes are usually selected to be as simple as possible, such as
the 2-state schemes for the bound and unbound states of enzymes or receptors
and the open and closed states of ion channels. 
Nevertheless, they can also be rather sophisticated (e.g., 8-state
inositol trisphosphate receptors \cite{Fall}, the 10-state hemoglobin \cite{Blatz}, and the 56-state chloride channels \cite{Blatz}).
The selection of kinetic schemes is mainly determined by the desired accuracy and the measurable quantities \cite{Hille,Keizer}.
Since a low-dimensional scheme can usually be contracted from higher-dimensional ones,
there exists a cascade of hierarchical Markovian network models suitable for 
describing the time evolution of the populations of a macromolecule's functional states \cite{Noe}.
These networks are anticipated to have indistinguishable kinetics,
exhibiting identical mean trajectories after being projected to the low-dimensional network space.
However, models with indistinguishable means do not necessarily have indistinguishable fluctuations.
A question that arises is that which schemes will give more relevant fluctuations to a real system and under which conditions unique fluctuation features can be obtained 
from different levels of contracted schemes?
These issues are essential for the reliability of various biological properties derived in terms of the fluctuations of a selected kinetic scheme, such as chemoreception \cite{Bialek_1,Wolde}, membrane conductance \cite{Chen5}, and ion channel density \cite{Sigworth}. 

The inter-network fluctuation relations arise from a comparison between different coarse-grained dynamical systems.
It resembles the comparison between different rate equations in the lumping analysis,
widely used in systems biology and general chemical engineering
\cite{Wei_1,Li_1,Toth,Okino,Gorban}. 
A central issue in that analysis is finding the lumping conditions for eliminating unimportant events
or time scales in a large network, of typically over $10^{4}$ species in systems biology, to reduce its complexity \cite{Liao}.
Interestingly, this contraction is mathematically analogous to merging experimentally indistinguishable states to obtain
simple transition networks for the conformational change of a macromolecule.
For instance, the Hodgkin-Huxley potassium ion channel has $16$ configurations depending on whether its individual four gates are open or closed \cite{Hille}.
However, this channel is often regarded as a 2-state system,
described by whether or not ions can pass through it in a patch-clamp recording.
The contraction from a 16-state to a 2-state model is
because the gating current recording is incapable of resolving the detailed structure of the channel configuration.
In terms of lumping analysis, this contraction is an approximate lumping \cite{Wei_2}.

Despite that correspondence, the original lumping analysis focuses on the relations between mean dynamics and is not concerned with fluctuations.
To extract this stochastic component, we generalize the lumping theory from original rate equations (RE) to chemical master equations (CME) and stochastic differential equations (SDE) and study kinetically equivalent (KE) and thermodynamically equivalent (TE) hierarchical kinetic schemes, under intrinsic and extrinsic noises.
The results go beyond the conventional assumption of ``fast relaxations" and contribute to our understanding of why a kinetic system can be contracted.
In the case of extrinsic noise, different kinetic schemes can give different fluctuations even when their average trajectories are the same.
This opens a possibility of identifying a correct kinetic model by observing fluctuations.
Notably, lumping conditions here are used for generating complex KE or TE networks from simple networks, in opposite to their original goal of reducing complex networks to simple networks.
Furthermore, for the conformational change of macromolecules discussed below, it is sufficient to focus on linear REs and linear lumping transformations.

\section{Lumping rate equations}
Let system $A$ be an $n$-dimensional kinetic scheme 
described by the linear RE,
\begin{equation}\label{}
  \frac{d{\bf N}}{dt}={\bf MN}\;\;
  \mbox{      or     }\;\; \frac{dN_i}{dt}=\sum_{j=1}^n k_{ji}N_j-k_{ij}N_i,
\end{equation}
where $N_i$ is the population of the $i$-th state and may represent the mean dynamics of some stochastic processes discussed later, ${\bf M}$ denotes the matrix of rate constants $k_{ij}$ from states $i$ to $j$, with $k_{ii}\equiv 0$,
and ${\bf N}\equiv[N_1,N_2,...,N_n]^T$ represents a state vector, in which the superscript $T$ stands for
the transpose of a vector.
If ${\bf U}$ is an $n'\times n$ full rank lumping matrix ($n'<n$), ${\bf N}$ can be contracted into
an $n'$-dimensional vector ${\bf N'}=[N'_1,N'_2,...,N'_{n'}]^T$ via
\begin{equation}\label{}
    {\bf N'}={\bf U}{\bf N},
\end{equation}
which is the state vector of some reduced system $A'$.
If each column of ${\bf U}$ is a standard unit vector, ${\bf U}$ denotes a proper lumping (see the example in S1 \cite{Supplemental}).
Since all lumpings in the following discussions are ``proper," this term will be neglected below.
The RE which ${\bf N'}$ satisfies is generally an integral-differential equation
with a memory kernel \cite{Keizer}.
If that kernel vanishes, the RE has a simple autonomous form as (1),
\begin{equation}\label{}
  \frac{d{\bf N'}}{dt}={\bf M'N'}\;\;
  \mbox{      or     }\;\; \frac{dN'_a}{dt}=\sum_{b=1}^n k'_{ba}N'_b-k'_{ab}N'_a,
\end{equation}
with $k_{aa}'\equiv 0$, and network $A$  is called ``exactly lumpable.''
Exact lumping makes the contracted system of an autonomous system
again autonomous, self-contained, and not having a memory kernel.
If the memory kernel does not vanish but is small, $A$ is called ``approximately lumpable,''
which has a broad practical application \cite{Wei_2}.
Exact lumping is the limiting case of all approximate lumpings when the memory effect tends to zero.
Equations (2) and (3) together constitute
the KE condition between $A$ and $A'$, or the condition for which $A$ can be exactly lumped into $A'$.
Notice that (2) alone is insufficient for this condition,
because any ${\bf U}$ can lump ${\bf N}$ into some ${\bf N'}$, which is not necessarily self-contained.

Quantitatively, the KE condition between $A$ and $A'$ can be expressed by their rate constant matrices
\begin{equation}\label{}
{\bf UM}={\bf M'U},
\end{equation}
which implies ${\bf U}e^{{\bf M}t}=e^{{\bf M'}t}{\bf U}$ \cite{Wei_1}.
When ${\bf U}$ is used to lump $A$ into $A'$, the $n$ states in $A$ are first partitioned into $n'$
sets $S_a$, with $a = 1, ..., n'$, by the row vectors of ${\bf U}$ (see S1 \cite{Supplemental}). 
Then, all states in $S_a$ are merged as the state $a$ in $A'$ and termed ``the internal states" of $a$.
Using the same procedure to merge all states in $S_a$ on both sides of (1),
one obtains the KE condition in terms of rate constants
\begin{equation}
k'_{ab}=\sum_{j\in S_b}k_{ij},
\end{equation}
for any $a$, $b\in\{1,2,...,n'\}$ with $a\neq b$ and any $i\in S_a$,
in analogy to that known for finite Markov chains \cite{Kemeny}.
Notice that the KE condition is fulfilled only when (5) is satisfied for all $i\in S_a$.
In brief, the KE condition can be expressed as (4) or (5), or equivalently as (2) together with (3).

Since (5) does not demand fast relaxations between the internal states in $S_a$,
the existence of fast variables or large $k_{ij}$ is not the prerequisite for exact lumpability.
However, lumping analysis can also eliminate fast variables,
as the quasi-equilibrium or quasi-steady-state approximations do \cite{Okino, Liao}.
Given a ${\bf U}$, whether $A$ described by (1) can be exactly lumped into $A'$ by ${\bf U}$
is decided by whether $A'$ has an autonomous RE (3), as discussed above.
If two autonomous $A$ and $A'$ are given first instead, whether $A$ can be lumped into $A'$ is decided by whether
some ${\bf U}$ can be found to connect them by (5).
If such ${\bf U}$ exists, ${\bf N'}$ of $A'$ and ${\bf N}$ of $A$ are indistinguishable,
in that the trajectories ${\bf N'}$ and ${\bf UN}$ are identical.

\section{Lumping master equations}
To extract the fluctuation relations of intrinsic noises between
hierarchical networks, we extend the lumping analysis from the RE (1) to its CME.
Suppose a macromolecule has $n$ conformational states whose
transition network $A$ obeys the kinetic equation (1).
If a system consists of $N$ macromolecules, its CME \cite{Hill_2,Chen_1},
\begin{eqnarray}\label{}
\frac{d{\bf P}}{dt}&=&{\bf LP} \mbox{  or }\\
\frac{dP_{\bf \tilde{N}}(t)}{dt}&=&\sum_{i,j=1}^nk_{ij}\left[(\tilde{N}_i+1)P_{{\bf \tilde{N}}-\boldsymbol{\omega}_{ij}}(t)-\tilde{N}_iP_{\bf \tilde{N}}(t)\right], \nonumber
\end{eqnarray}
describes the evolution of the joint probability $P_{\bf \tilde{N}}(t)$ of finding the state vector
${\bf \tilde{N}}\equiv[\tilde{N}_1,\tilde{N}_2,...,\tilde{N}_n]^T$ at time $t$,
where $\tilde{N}_i\geq 0$ is the number of macromolecules in the $i$-th state and $\sum_{i=1}^n \tilde{N}_i= N$.
Therein, ${\bf \tilde{N}}$ is related to the ${\bf N}$ in (1) by
$\sum_{{\bf \tilde{N}}}  \tilde{N}_i P_{\bf \tilde{N}}(t)=N_i$,
where the sum runs over all accessible ${\bf \tilde{N}}$.
The vector $\boldsymbol{\omega}_{ij}$ has values $-1$ and $+1$ in its $i$-th and $j$-th components, respectively,
and $0$ elsewhere.
It stands for the change of molecule numbers in different states during the reaction
shifting one molecule from $i$ to $j$.
Notice that ${\bf P}$ is a vector whose ``${\bf \tilde{N}}$-th" component is the probability $P_{\bf \tilde{N}}(t)$,
just as ${\bf N}$ in (1) is a vector whose $i$-th component is $N_i$.

For each lumping matrix ${\bf U}$, which contracts ${\bf N}$ of $A$ into ${\bf N'}={\bf U}{\bf N}$ of $A'$,
there exists an associated lumping operator ${\bf \widehat{U}}$,
which contracts ${\bf P}$ into a reduced vector
\begin{equation}\label{}
{\bf P'}={\bf \widehat{U}}{\bf P},
\end{equation}
whose ${\bf \tilde{N}'}$-th component is (see S2 \cite{Supplemental})
\begin{equation}\label{}
P'_{\bf \tilde{N}'}(t)=\sum_{{\bf \tilde{N}}}P_{\bf \tilde{N}}(t)\prod_{c=1}^{n'}
\delta\left(\tilde{N}_c'-\sum_{k\in S_c}\tilde{N}_k\right),
\end{equation}
where sets $S_c$ are partitioned by ${\bf U}$ as explained in the text that follows (4)
and $\delta(X'-X)$ is a Kronecker delta whose value is one when $X'=X$ and zero elsewhere.
If ${\bf U}$ is arbitrary, ${\bf N'}$ does not necessarily obey a simple RE as (3) and
$P'_{\bf \tilde{N}'}(t)$ does not necessarily satisfy any CME of the same form as (6).
However, if ${\bf U}$ can exactly lump $A$ into $A'$, ${\bf N'}$ does follow (3) and
$P'_{\bf \tilde{N}'}(t)$ indeed obeys a simple lumped CME
\begin{eqnarray}\label{}
\frac{d{\bf P'}}{dt}&=&{\bf L'P'} \mbox{  or }  \\
\frac{dP'_{\bf \tilde{N}'}(t)}{dt}&=&\sum_{a,b=1}^{n'} k'_{ab}\left[(\tilde{N}'_a+1)P'_{{\bf \tilde{N}'}-\boldsymbol{\omega}_{ab}}(t)-\tilde{N}'_aP'_{\bf \tilde{N}'}(t)\right], \nonumber
\end{eqnarray}
which turns out to be the CME of $A'$ (see S2 \cite{Supplemental}).
Alternatively, suppose the REs of $A$ and $A'$ are (1) and (3) and some ${\bf U}$ can exactly lump $A$ into $A'$ through (2).
Then their ${\bf P}$ and ${\bf P'}$ in (6) and (9) are related by (7) and thus indistinguishable from each other,
which is the exact lumpability in terms of joint probabilities.
Just as (2) and (3) form the KE condition between two REs, (7) and (9) constitute the KE condition on the level of CME.
With the same argument as for (4), the lumping condition for the CME is
\begin{equation}\label{}
  {\bf \widehat{U}}{\bf L}={\bf L'}{\bf \widehat{U}}.
\end{equation}
Notably, the exactly lumped CME (9) via the KE condition is distinct from the reduced CME entirely based
on the time scale separation \cite{Roussel_2}.

The above argument indicates that the exact lumpability in RE (1) implies the exact lumpability in its CME (6)
and vice versa (see S2 \cite{Supplemental}).
Therefore, the KE condition is a rather strong condition for systems under intrinsic noises.
It not only conveys the original meaning of identical first moments, ${\bf UN}$ and ${\bf N'}$,
but also the identities of all other moments, owing to the identity of probabilities, ${\bf \widehat{U}}{\bf P}={\bf P'}$, (see (2.13) in S2 \cite{Supplemental}).
Physically it indicates that experimentally measured fluctuations cannot be used for
judging whether a state has internal states,
if the fluctuations are caused by small numbers of macromolecules. 

Among all moments, of special interest are the indistinguishable second moments,
\begin{equation}\label{}
\boldsymbol{\tilde{\sigma}'}={\bf U}\boldsymbol{\tilde{\sigma}}{\bf U}^T,
\end{equation}
where
$\tilde{\sigma}_{ij}\equiv\langle\delta N_i\delta N_j\rangle$
($\tilde{\sigma}'_{ab}\equiv\langle \delta N'_a\delta N'_b\rangle$)
is an average over the probability $P_{\bf \tilde{N}}(t)$ ($P'_{\bf \tilde{N}'}(t)$)
and $\delta N_i=\tilde{N}_i-N_i$ ($\delta N_a'=\tilde{N}_a'-N_a'$) is the fluctuation
around the mean $N_i$ of $\tilde{N}_i$ ($N'_a$ of $\tilde{N}_a'$) defined in (6).
In Fig. 1, the indistinguishable
variances,
$\tilde{\sigma}_{ii}$ and ${\bf \tilde{\sigma}'}_{aa}$, induced by the intrinsic noises of two KE networks,
are numerically confirmed.

Besides the strict KE condition, a kinetic scheme may be selected merely because it is TE to the real system
\cite{Wales}.
If two kinetic networks $A$ and $A'$ are TE to each other,
the stationary states ${\bf N'}^s$ and ${\bf N}^s$ of their REs are related
by ${\bf N'}^s={\bf UN}^s$ via some lumping matrix ${\bf U}$.
The stationary solution of the CME is the multinomial distribution,
\begin{equation}\label{}
P^s_{\bf \tilde{N}}=\frac{N!}{\prod_{i=1}^n\tilde{N}_i!}\prod_{j=1}^n\left(\frac{N_j^s}{N}\right)^{\tilde{N}_j},
\end{equation}
where $N_i^s$ is the $i$-th component of ${\bf N}^s$ \cite{Hill_2}.
Let ${\bf P}^s$ be the vector whose ${\bf \tilde{N}}$-th component is the ${P}^s_{\bf \tilde{N}}$ of $A$ and
${\bf P'}^s$ be the vector whose ${\bf \tilde{N}'}$-th component is the ${P'}^s_{\bf \tilde{N}'}$ of a TE system $A'$
of $A$.
One can show that ${\bf P}^s$ and ${\bf P'}^s$ are related by (see S3 \cite{Supplemental})
\begin{equation}\label{}
{\bf P'}^s={\bf \widehat{U}}{\bf P}^s
\end{equation}
irrespective of whether $A'$ is KE to $A$ or not. 
More precisely, (13) is sufficient and necessary for the TE condition ${\bf N'}^s={\bf UN}^s$,
or is the TE condition on the level of stationary joint probability (see S3 \cite{Supplemental}).
While under the KE condition the contracted probability $P'_{\bf \tilde{N}'}(t)$ must satisfy
(9) at any $t$, under the TE condition it must only obey the form (12)
at $t=\infty$.

An arbitrary network $A$ does not always have a reduced KE system.
However, it usually has infinitely many reduced TE systems $A'$'s, which are TE to one another.
An interesting indication from (7) and (13) is that
if $A$ and $A'$ are TE, but not KE, to each other, their initially distinguishable ${\bf P}$ and ${\bf P'}$
will become indistinguishable as $t\rightarrow\infty$, irrespective of which ${\bf U}$ is
used to contract $A$ to $A'$ (Fig. 2).
Therefore, the lumpability between the probabilities of TE systems is similar to the Lyapunov function
for quantifying entropy production, where Kullback-Leibler divergence may be a proper lumpability measure.

\section{Lumping stochastic differential equations}

Another frequently used approach for exploring fluctuations is the SDE,
\begin{equation}\label{}
\frac{d{\bf {\hat N}}}{dt}={\bf M}{\bf {\hat N}}+{\bf f},
\end{equation}
where ${\bf {\hat N}}={\bf N}+\delta {\bf N}$ is a real-valued random variable with the fluctuations $\delta {\bf N}$
about the ensemble mean ${\bf N}$, which satisfies a deterministic equation as (1).
Here, ${\bf f}$ is a Gaussian white noise with $\langle {\bf f}(t)\rangle={\bf 0}$,
$\langle {\bf f}(t'){\bf f}^T(t)\rangle={\bf \Gamma}\delta(t-t')$,
and $\langle {\bf f}(t'){\bf \hat{N}}^T(t)\rangle={\bf 0}$ for $t<t'$,
where the covariance matrix ${\bf \Gamma}$ is symmetric, positive semi-definite, and generally time-dependent.
The solution of (14), ${\bf \hat{N}}(t)=e^{{\bf M}t}{\bf \hat{N}}(0)+\int_0^te^{{\bf M}\tau }{\bf f}(t-\tau )\,d\tau$,
is also a Gaussian random variable.
The conditional covariance of $\delta{\bf N}$ is
$\boldsymbol{\sigma}\equiv\left\langle\delta{\bf N}\delta{\bf N}^T\right\rangle
=\int_0^t e^{{\bf M}\tau }{\bf \Gamma} \left(e^{{\bf M}\tau }\right)^T\,d\tau$,
which is symmetric and has the time derivative
$d{\boldsymbol{\sigma}}/dt={\bf M}{\boldsymbol{\sigma}}+{\boldsymbol{\sigma}}{\bf M}^T
+{\bf \Gamma}$.
This equation is reduced to the fluctuation-dissipation theorem (FDT) when the system reaches equilibrium
as $t\rightarrow\infty$, where $d{\boldsymbol{\sigma}}/dt$ vanishes \cite{Keizer}.
If ${\bf f}$ represents an intrinsic noise, $\boldsymbol{\sigma}$ will be the
$\boldsymbol{\tilde{\sigma}}$ in (11), when the system is close to the thermodynamic limit.
Together with the given ${\bf M}$ it uniquely determines ${\bf \Gamma}$ via the FDT.
The ${\bf \Gamma}$ in the chemical Langevin equation in Ref. \cite{Gillespie1} belongs to this category.
If ${\bf f}$ is an extrinsic noise, $\boldsymbol{\sigma}$ and ${\bf \Gamma}$ can be freely tuned as long as they
comply with the FDT.

Let $A$ and $A'$ be two network models approaching a real system,
where $A$ is described by (14) and $A'$ satisfies
\begin{equation}\label{}
  \frac{d{\bf \hat{N}'}}{dt}={\bf M'}{\bf \hat{N}'}+{\bf f'}.
\end{equation}
Here ${\bf {\hat N}'}={\bf N'}+\delta {\bf N'}$ and ${\bf f'}$ has statistical properties analogous to ${\bf f}$.
If $A$ and $A'$ are KE to each other, they are connected by some ${\bf U}$ via ${\bf N'}={\bf UN}$
(notably not ${\bf \hat{N}'}={\bf U\hat{N}}$).
The covariance of the fluctuations of ${\bf U\hat{N}}$ is
${\bf U}\boldsymbol{\sigma}{\bf U}^T=\left\langle{\bf U}\,\delta{\bf N}\,\delta{\bf N}^T{\bf U}^T\right\rangle
=\int_0^t {\bf U}e^{{\bf M}\tau }{\bf \Gamma} \left(e^{{\bf M}\tau }\right)^T{\bf U}^T\,d\tau
=\int_0^t e^{{\bf M'}\tau }{\bf U}{\bf \Gamma}{\bf U}^T \left(e^{{\bf M'}\tau }\right)^T\,d\tau$,
where the exchange relation implied by (4) has been used to obtain the last equality.
Since ${\bf U}\boldsymbol{\sigma}{\bf U}^T$ is indistinguishable from $\boldsymbol{\sigma}$,
the distinguishiability between the covariances $\boldsymbol{\sigma'}$ and $\boldsymbol{\sigma}$
of two KE systems $A'$ and $A$ can be determined by the difference
\begin{equation}\label{}
\boldsymbol{\sigma}_{\rm diff}\equiv\boldsymbol{\sigma'}-{\bf U}\boldsymbol{\sigma}{\bf U}^T
=\int_0^t e^{{\bf M'}\tau }{\bf \Gamma}_{\rm diff}\left(e^{{\bf M'}\tau }\right)^T\,d\tau,
\end{equation}
where ${\bf \Gamma}_{\rm diff}\equiv{\bf \Gamma'}-{\bf U}{\bf \Gamma}{\bf U}^T$ is a time-dependent
symmetric matrix.
While (16) tells us that ${\bf \Gamma}_{\rm diff}={\bf 0}$ implies $\boldsymbol{\sigma}_{\rm diff}={\bf 0}$,
its time derivative,
$d\boldsymbol{\sigma}_{\rm diff}/dt=e^{{\bf M'}t}{\bf \Gamma}_{\rm diff}\left(e^{{\bf M'}t}\right)^T$,
implies the opposite, since $e^{{\bf M'}t}$ is an invertible matrix.
Thus, $\boldsymbol{\sigma}_{\rm diff}={\bf 0}$ if and only if
\begin{equation}\label{}
{\bf \Gamma}_{\rm diff}={\bf 0}\mbox{, or equivalently  } {\bf \Gamma'}={\bf U}{\bf \Gamma}{\bf U}^T.
\end{equation}
This relation was already known for ${\bf U}$ replaced by invertible transformations  ((8.2.39) in Ref. \cite{Keizer}),
for which the argument is more straightforward than that for (17).

Relation ${\bf \Gamma}_{\rm diff}={\bf 0}$ in (17) is a weak condition, under which $A$ and $A'$ have only ``statistically" indistinguishable ${\bf \hat{N}}$ and ${\bf \hat{N}'}$.
A plausible stronger condition is
\begin{equation}\label{}
{\bf f'}={\bf U}{\bf f},
\end{equation}
which fulfills (17) and generates indistinguishable individual stochastic trajectories ${\bf \hat{N}}$ and ${\bf \hat{N}'}$.
Both (17) and (18) lead to indistinguishable covariances and variances of fluctuations of  ${\bf \hat{N}}$ and ${\bf \hat{N}'}$.
Together with the indistinguishable means of the KE condition, it yields the indistinguishable
Gaussian distributions of ${\bf \hat{N}}$ and ${\bf \hat{N}'}$.

Although (17) shows that $\boldsymbol{\sigma}_{\rm diff}={\bf 0}$ if and only if
${\bf \Gamma}_{\rm diff}={\bf 0}$, it
does not reveal whether two KE systems should have $\boldsymbol{\sigma}_{\rm diff}={\bf 0}$ or not.
For intrinsic noises, the indistinguishable covariances in (11) from the CME approach lead to the expectation that
$\boldsymbol{\sigma}_{\rm diff}={\bf 0}$ in the SDE approach, because the SDE can describe
CME fluctuations near the thermodynamic limit.
According to (17), this expectation would be true
if ${\bf \Gamma}_{\rm diff}={\bf 0}$,
which indeed can be proved (see $\Gamma_{ij}$ of ion channels below and S4 \cite{Supplemental}).
For extrinsic noises, ${\bf \Gamma}$ is not decided by ${\bf N}$
and distinct ${\bf \Gamma}$'s will generate different fluctuations.
Let ${\bf V}_{\rm diff}$ be a variance matrix whose diagonal terms
are the same as those of $\boldsymbol{\sigma}_{\rm diff}$ and zero elsewhere.
For two KE systems $A$ and $A'$, (16) implies the simple ordering rule for the variances of their state fluctuations
at any $t$:
\begin{equation}\label{}
  {\bf \Gamma}_{\rm diff}\geq {\bf 0}  (\leq {\bf 0}, = {\bf 0}) \Rightarrow {\bf V}_{\rm diff}\geq {\bf 0} (\leq {\bf 0}, = {\bf 0}),
\end{equation}
where $\geq {\bf 0}$ ($\leq {\bf 0}$) and $={\bf 0}$ stand for positive (negative) semi-definite and null matrices, respectively.

In practice, which of (17), (18), and (19) is the correct relation between two KE models $A$ and $A'$ of
a real macromolecule depends on what we study.
For intrinsic noises, the ${\bf \Gamma}$ and ${\bf \Gamma'}$ of $A$ and $A'$
can be analytically derived and must be related by (17).
For extrinsic noises, if $A$ and $A'$ are to approach the same experimental data,
their covariances should obey (18).
If $A$ and $A'$ are to approach two individually measured experimental data of the same macromolecule,
their fluctuations may have diverse orderings (19), because environmental noises in different
experiments are likely different.
Yet, if the noises are statistically the same, the covariance relation is (17), as for intrinsic noises.

Experimentally, fluctuations have been measured to predict the ion channels density, e.g., in nerve fibers of {\it Rana pipiens} \cite{Sigworth}.
To model this experiment with SDE (14), one considers a variety of channels,
each of which can stochastically transit between $n$ conformational states, with transition probabilities
given by the rate constants in (1).
According to the canonical theory \cite{Keizer} or the linear noise approximation \cite{van_Kampen_1},
the stochastic force ${\bf f}$ in (14) has the covariance
$\Gamma_{ij}=\sum_{k=1}^n(k_{ki}N_k+k_{ik}N_i)\delta_{ij}-(k_{ij}N_i+k_{ji}N_j)$,
where $N_i$ is the probability of finding a channel in the $i$-th state and
$\delta_{ij}$ denotes the Kronecker delta.
This covariance depends on the evolution of the mean value $N_i$ and thus varies with time.
If the channel is modeled by a two-state (open/closed) system,
the $(1,1)$ entry of its equilibrium covariance \cite{Fall}, $\Gamma_{11}^e=k_{12}N_1^e+k_{21}N_2^e$,
complies with Onsager's statistical theory of equilibrium ensembles \cite{Keizer}. 
If the channel is modeled by two KE systems of different dimensions with the same form as $\Gamma_{ij}$,
they fulfill $\boldsymbol{\Gamma}_{\rm diff}={\bf 0}$ in (17) (see S4 \cite{Supplemental})
and then $\boldsymbol{\sigma}_{\rm diff}={\bf V}_{\rm diff}={\bf 0}$.
Therefore, the indistinguishability $\boldsymbol{\sigma}_{\rm diff}={\bf 0}$ from the Gaussian probability
in the SDE approach coincides with the indistinguishability (11) from the joint probability in the CME approach.

\section{Conclusion}

Theoretically we generalized the lumping theory from deterministic dynamics to stochastic processes.
It allows us to compare stochastic properties between hierarchical networks,
such as networks of small systems, which are sensitive to
external noises, or large networks whose species contain small number of copies.
In applications, we introduced lumping techniques from systems biology to molecular biology
to explore the fluctuation relations of experimentally indistinguishable kinetic schemes of biomolecules
and the legitimacy of estimating macromolecular fluctuations by low-dimensional schemes.
These findings are a kind of contractions beyond the widely discussed ones based on ``fast relaxations" and are useful for extracting correct kinetic models by observing extrinsic noise induced fluctuations.
The analytical results derived from exact lumping here provide limiting properties
for networks connected by all kinds of approximate lumping conditions.
They further give insights into more general fluctuation relations in other contraction theories,
which usually utilize similar block-triangular matrices to reduce systems
\cite{Liao}, such as Keizer's memoryless contraction \cite{Keizer}
and hierarchical Volterra equations in the Zwanzig-Mori formalism \cite{Berne}.
For further study, one may take into account more subtle issues, such as the approximate lumping for non-Markovian networks \cite{Wei_2} and the deformation of hidden complexity of free energy surfaces \cite{Krivov}.

\section*{Acknowledgments}
We thank Tetsuya J. Kobayashi, Jun Ohkubo, Jung-Hsin Lin, and Lee-Wei Yang
for useful discussions, the National Center for Theoretical Sciences at Taiwan for its supports, and the support of the Ministry of Science and Technology of Taiwan through Grant No. NSC 102-2112-M-009-012.


\newpage

\begin{figure}
\centerline{\includegraphics[width=0.82\textwidth]{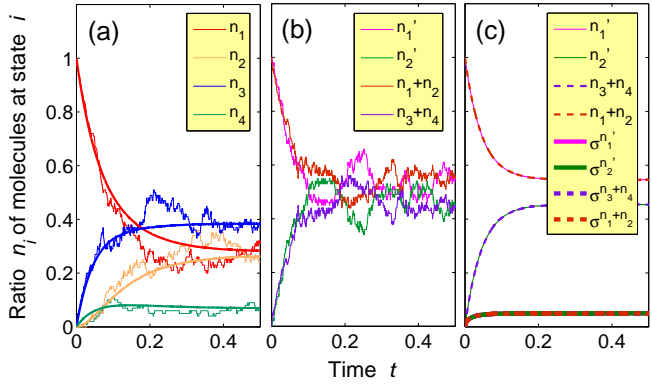}}
\caption{A system consists of $N=10^2$ macromolecules, each of which can be regarded as a four-state system $A$ with
$[k_{13},k_{31},k_{14},k_{41},k_{23},k_{32},k_{24},k_{42}]=[8,6,2,7,9,6,1,5]$ and $k_{12}=k_{21}=k_{34}=k_{43}=0$
or a two-state system $A'$ with $[k'_{12},k'_{21}]=[10,12]$, satisfying the KE condition (5).
The inter-state transitions are stochastic and follow the probabilities assigned by the rate constants in RE (1).
(a) The four erratic curves are the stochastic trajectories of the ratio, $n_i\equiv\tilde{N}_i/N$,
of the molecules in the $i$-th state, with $i=1, ..., 4$, calculated by this Markov chain to simulate the results of the CME.
Averaging over $10^4$ realizations, the intrinsic fluctuations tend to zero and the
four erratic curves become four smooth curves.
(b) The dynamics of $n_1+n_2$ and $n_3+n_4$
recorded from a single realization of $A$ are only roughly close to those of $n'_1$ and $n'_2$ of its KE system $A'$.
(c) Averaging over $10^4$ realizations, $n_1+n_2$ and $n_3+n_4$
precisely approach $n'_1$ and $n'_2$, respectively.
The coincidence in the mean dynamics leads to the coincidence in their variances,
as shown by the four overlapped thick lines at the bottom.
} \label{}
\end{figure}

\begin{figure}
\centerline{\includegraphics[width=0.49\textwidth]{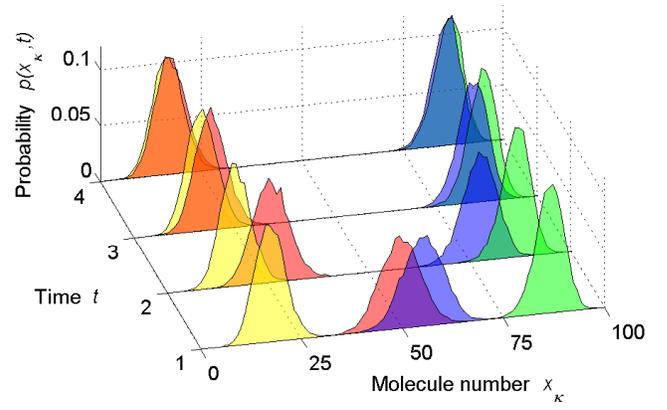}}
\caption{A system consists of $10^2$ identical macromolecules, each described by a three-state
transition network $A$ with the rate constants $[k_{12},k_{21},k_{23},k_{32},k_{31},k_{13}]=[0.07,1,0.5,9,0.4,30]$.
Network $A$ can be contracted into a two-dimensional TE network $A'$ ($A''$) by merging states
$1$ and $2$ ($1$ and $3$) of $A$ into state $1'$ of $A'$ ($A''$) and renaming state $3$ ($2$) of $A$ as
state $2'$ of $A'$ ($A''$).
The probability, $p(x_\kappa,t)$, of finding $x_\kappa$ macromolecules in the
$\kappa$-th state at time $t$ is estimated
by counting the frequency of that event when the system evolves $10^2$ times.
Two initially distinct distributions $p(\tilde{N}_1+\tilde{N}_2,t)$ of $A$ (blue) and $p(\tilde{N}_1',t)$ of $A'$ (green),
as well as $p(\tilde{N}_1+\tilde{N}_3,t)$ of $A$ (red) and $p(\tilde{N}_1'',t)$ of $A''$ (yellow),
approach each other as $t\rightarrow\infty$.
This example demonstrates the increasing lumpability between the probabilities of two TE networks,
as indicated by (7) and (13), in terms of the marginal probability $p(x_\kappa,t)$.
}\label{}
\end{figure}

\end{document}